\documentclass[12pt]{article}
\usepackage{graphicx}
\usepackage{bm}
\newcommand\dpar[2]{\frac{\partial#1}{\partial#2}}
\def\be{\begin{equation}}
\def\ee{\end{equation}}
\def\ba{\begin{eqnarray}}
\def\ea{\end{eqnarray}}
\def\half{\frac{1}{2}}
\def\Idep{I_{dep}}
\def\qdep{q_{dep}}
\def\D{\Delta}
\def\Drho{\Delta_{,\rho}}
\def\sf{\sqrt{f}}
\def\sfz{\sqrt{f_0}}
\def\X{{\bf X}}
\def\eps{\epsilon}
\def\Jinf{J_{\infty}}
\def\qmin{q_{\min}}
\begin{document}
\begin{titlepage}
\title{\Large \bf Gapless superconductivity and string theory}
\author{Sergei Khlebnikov\thanks{\tt skhleb@purdue.edu} \\
\textit{Department of Physics and Astronomy} \\
\textit{Purdue University, West Lafayette, IN 47907}}
\date{}
\maketitle
\begin{abstract}
Coexistence of superconducting and normal components in nanowires at currents below 
the critical (a ``mixed'' state) would have important consequences for the nature and 
range of potential applications of these systems. 
For clean samples, it represents a genuine
interaction effect, not seen in the mean-field theory. 
Here we consider properties 
of such a state in the gravity dual of a strongly coupled superconductor constructed from
D3 and D5 branes. We find numerically uniform gapless solutions containing both components
but argue that they are unstable against phase separation, as their free energies 
are not convex. We speculate on the possible nature of the resulting non-uniform sate 
(``emulsion'') and draw analogies between that state and the familiar mixed state of a 
type II superconductor in a magnetic field. 
\end{abstract}
\end{titlepage}
\setcounter{page}{2}
\tableofcontents
\section{Introduction}
Many proposed applications of nanoscale superconductors require understanding of how 
these systems behave under currents close to the critical. For instance, in designing
superconducting qubits, it is essential to know how to use current to suppress
the potential barrier separating the basis states. 

The best studied example of nanoscale superconductor is a point-like weak link---a 
Josephson junction (JJ). It can often be described by a single-degree of
freedom $\theta$---the phase difference between the leads---subject to a ``tilted washboard'' 
potential
\be
V(\theta) = - V_0 \cos\theta - I \theta\, .
\label{wash}
\ee
Here $I$ is the electric current in units
of $2e$; $e$ is the electron charge. For {\em static} $\theta$ (when $V$ equals the total
energy of the system), variation of (\ref{wash}) 
produces the equation $V_0 \sin \theta = I$,
showing that the current $I$ is due
to a gradient of the phase, that is, $I$ is entirely a supercurrent: $I= I_s$. 
For a time-dependent $\theta$, however, the current contains
both superconducting and normal components. Thus, in general
\be
I = I_s + I_n \, .
\label{I}
\ee
The normal component $I_n$ includes not only the normal current through the JJ itself
(due, for instance, to thermal quasiparticles)
but also currents through various external resistors (``shunts'') connected 
in parallel to it. This is because all such currents couple to $I_s$ via voltage
fluctuations, which are proportional to the time derivative of $\theta$. 

For any $-V_0 < I < V_0$, the potential (\ref{wash}) has infinitely many minima, 
equally spaced
by $2\pi$, and important fluctuations are those that take the system from one minimum to 
the next. These are known as phase slips. Each phase slip generates a voltage spike in
the external circuit. If such spikes occur at a non-negligible rate, at a finite
$I$ they will produce a nonzero time-averaged voltage, i.e., a finite resistance.

Recently, a number of experimental techniques have been developed for synthesizing
systems in which superconductivity is one-dimensional---superconducting (SC) nanowires. 
These techniques,
described in the books \cite{Bezryadin,Altomare&Chang}, 
result in wires of uniform thickness with linear
cross-sectional dimensions of a few nanometers. For such thin wires, one can assume 
that SC properties (e.g., the supercurrent density) depend only on the lengthwise
direction, 
even though the electron density of states still retains the 3d character.
These novel systems promise a potentially new class of devices for control 
of superconductivity by current.

Even though we do not expect the model (\ref{wash}) to apply literally to the case of
wires, some of the notions discussed above do carry over. The phase of the order parameter,
$\phi(x,t)$, is now a function of the coordinate $x$ along the wire and time $t$.
The supercurrent is proportional to the gradient of the phase: 
$I_s \propto \nabla \phi$. For states where $\phi$ is a continuous function of $x$, we
can identify the winding number as
\be
W(t) = \frac{1}{2\pi} \left[ \phi(L,t) - \phi(0,t) \right] 
\label{W}
\ee
($L$ is the length of the wire)
and a phase slip as an event that changes $W$ by $\pm 1$. There is 
a novel aspect to a phase slip in a wire (as opposed to the case of JJ), which has to do
with the continuity of $\phi$. Namely, the process now occurs locally, at some
point $x$, where the order parameter momentarily vanishes, allowing the phase
to unwind \cite{Little}. 

Similarly to the case of JJ, one can imagine a nanowire shunted by various
external impedances, resulting in
a normal current connected in parallel to the superconducting one, for the total
given by the same eq.~(\ref{I}). In this paper, however, we wish to consider 
the possibility 
of an {\em intrinsic} resistive effect, namely, a normal component that is formed 
in the wire itself (and, unlike thermal quasiparticles, survives in the limit $T\to 0$).
Such a resistor will remain even in a wire effectively decoupled from
any external dissipative environment, for example, in a SC loop operated via inductive
coupling to a coil.

For a nanowire, one can consider, at least theoretically, two extreme 
limits. One is the clean limit---a perfectly uniform wire without disorder; the
other is the dirty limit---a wire with strong disorder scattering. The second limit
is presumably more realistic, but the first is simpler and, as such, may be a useful
starting point. In this paper, we consider the clean limit exclusively. 
We also restrict ourselves to $T = 0$,
where the question of existence of an intrinsic normal 
component is in a sense the sharpest, although the method we propose
is applicable also at $T\neq 0$.

As in a JJ, the SC and normal components in a wire are coupled via phase slips.
The clean limit is momentum-conserving, so the momentum unwound by a phase slip 
from the supercurrent must be picked up, in its entirety, by the normal component.
The latter may in principle include small oscillations of the SC density (the plasma
waves \cite{Kulik,Mooij&Schon}), 
which, similarly to waves in a waveguide, are characterized by a finite impedance, but a
more detailed study shows that there is a quantum anomaly involved, and each phase slip 
produces, via level crossing, fermionic quasiparticles, in precisely the right number
to account for conservation of momentum \cite{Khlebnikov:2003}.

The requirement of quasiparticle production affects the energy balance in a phase slip: 
for the process to occur spontaneously, the energy unwound from
the supercurrent must be enough to offset the cost of the produced fermions.
In mean-field theory, the free energy unwound from a loop of length $L$ gives directly the
supercurrent:
\be
I_s = \frac{1}{2\pi} \dpar{F}{W} \, ,
\label{unw}
\ee
where $W$ is the winding number (\ref{W}). 
So, one may suppose that phase slips become more favorable at 
larger $I_s$. While that is true to a degree, a direct calculation 
shows that, within
mean-field theory, the energy $\partial F / \partial W$ is never large enough---that is
not until $I_s$ reaches the depairing current $\Idep$.\footnote{This conclusion holds 
rather generally, provided one neglects corrections suppressed by the ratio of the gap
to the energy scale 
of the band structure. It does not depend on 
Galilean invariance or other such special symmetries.} The question we wish to ask is
whether this conclusion is a mean-field artifact; in other words, whether a window
in which SC and normal components can coexist will open (below $\Idep$) once electron-electron 
interactions are fully taken into account.

One possible way to answer this question is to construct a superconductor from strings
and branes and go over to the strong-coupling (large $N$) limit, in which $N$ coincident
branes behave as a classical gravitating object \cite{Polchinski:1995mt}; such an alternative 
description of a quantum system is know as a gravity dual. 
A well-known example of gauge/gravity duality is the AdS/CFT correspondence 
\cite{Maldacena:1997re},
for which a ``holographic'' dictionary connecting the two sides of the duality
has been established 
\cite{Gubser:1998bc,Witten:1998qj}. Calculations using a gravity dual, however, are possible 
even in cases where a complete dictionary is not known, as long as one concentrates on those 
quantities that can in fact be unambiguously defined on the gravity side.
The quantity we are interested in here is the energy of the ground
state, $F(P_s)$, as a function of the momentum of the SC component, $P_s$, at
fixed total momentum $P$. The difference 
\be
P_n = P - P_s
\label{Pn}
\ee
can then be attributed to the normal component. This can be seen as a clean-limit
version of the formula (\ref{I}).\footnote{In our earlier paper 
\cite{Khlebnikov:2012yd}, current and momentum were 
spoken of largely interchangeably. Here, we aim to be more careful about the distinction.}
A ground state with both component present corresponds to a minimum of $F(P_s)$ for which
both $P_s$ and $P_n$ are nonzero.

A brane construction suitable for modeling a SC nanowire has been proposed in 
\cite{Khlebnikov:2012yd}. It is based on a system of D3 and D5 branes in type IIB
string theory.\footnote{Our approach is different from other 
holographic descriptions 
of superconductivity that have been proposed in the literature. It is distinct from
the one in
\cite{Gubser:2008px,Hartnoll:2008vx,Nakamura:2009tf} in that it does not use a bulk
U(1) gauge field. (States with nonzero supercurrent in the model of 
\cite{Gubser:2008px,Hartnoll:2008vx} have been considered in 
\cite{Basu:2008st,Herzog:2008he,Sonner:2010yx,Amado:2013aea}.) And our approach is distinct
from the brane construction
of \cite{Ammon:2008fc} in that it preserves the worldvolume gauge symmetry. 
We use the corresponding conserved charge to describe the linear momentum 
(quantized in units of the Fermi momentum $k_F$).} The key aspects of it are as follows:

(i) The setup contains a large number $N$ of coincident D3 branes and a D5 probe
intersecting them over a line. This breaks all supersymmetries.
The direction along the line, $x \equiv x^1$, corresponds to the direction
of the wire, and the two directions transverse to all branes, $x^8 + i x^9$, to the SC
order parameter. (ii) $N$ is identified with the number of occupied
channels (transverse wavefunctions) in the wire, 
$N = N_{ch}$.\footnote{$N_{ch}$ 
is proportional to the cross-sectional area of the wire and in practice is 
of order of a few thousand, for the thinnest wires available. The large value 
reflects the 3d character
of the electron density of states; this is in contrast to SC properties, 
which vary only along $x$.} (iii) Supercurrent corresponds to the D5 winding around the 
D3s as one moves along $x$. If the branes actually intersect, the low-energy
spectrum of 3/5 strings contains $N$ species of massless (1+1)-dimensional fermions
(both left and right movers). At low values of $P$, however, the intersection is unstable,
and the D5 moves a finite distance away from the D3s; this corresponds to a fully gapped,
supercurrent-only state. 

Supercurrent-only solutions have been found 
in \cite{Khlebnikov:2012yd}. The Chern-Simons 
term in the D5 action has the effect that the solution with
winding number $W$ carries $N W$ units of the D5 worldvolume charge. A phase slip
corresponds to the D5 crossing the D3s, with $W$ changing by one. Conservation of
charge then implies that $N$ fundamental strings, stretching between the D5 and
and the D3s, must be produced. This can be seen, on the one hand, as a version of the
Hanany-Witten effect \cite{Hanany:1996ie} in string theory
(creation of branes and strings at intersections) and, 
on the other, as a parallel to the requirement of quasiparticle production
noted earlier. This parallel 
allows one to identify the worldvolume charge with the supercurrent momentum $P_s$ (which,
for a supercurrent-only state,
is also the total momentum), as follows: 
\be
P_s = N W \, ,
\label{NW}
\ee
where by convention $P_s$ is in units of the Fermi momentum. 
Instead of $W$, we will often use the
winding number density
\be
q = 2\pi W / L \, ,
\label{q}
\ee
where an extra factor $2\pi$ is added for convenience.

In the leading large-$N$ limit, supercurrent-only solutions exist for all $q$, no matter
how large.
At $q$ above a certain $q_m$, however, phase slips become energetically favorable,
and the supercurrent-only state unstable. The instability has nothing to do with depairing.
Indeed, $q_m \sim 1/R$, where $R$ is the length scale of the D3 metric (the only length 
scale seen by classical gravity). Meanwhile, the value $q= \qdep$ corresponding to depairing 
is of order of the gap $\D$, i.e., of order $R$ in units of the string tension. One sees that
the ratio $\qdep / q_m$ scales to infinity in the large $N$ limit.\footnote{
This may explain why no depairing is seen in the calculation 
of \cite{Khlebnikov:2012yd}.} 

Instability of the supercurrent-only state at $q > q_m$ means that for
\be
P > P_m = N q_m L / 2\pi \,
\label{Pm}
\ee
not all of the total momentum $P$ in the true ground state is carried by supercurrent;
some must be carried by quasiparticles. On the other hand, we find that the
normal-only state is unstable
for any $P$. We conclude that, at $P$ exceeding the bound (\ref{Pm}), 
supercurrent and quasiparticles must coexist in some mixed state.

A priori, it is not clear what the nature of the mixed state is and, in particular, if 
it can be described by a uniform (in $x$) time-independent solution of dual 
gravity (as the supercurrent-only state could). 
Nevertheless, looking for such solutions
is a natural first step, and that is what we describe in this paper. We find that, for a given 
total $P$, a uniform time-independent solution exists for all $P_s$ between
the value at which the normal-only state first becomes unstable and the maximum
$P_s = P$. For these solutions, the D5 crosses the D3s' horizon; we argue their
existence by considering the near-horizon limit and by numerical evidence. We find,
however, that the free energy of such a solution is never a convex function of $P_s$.
This means that the uniform mixed state is unstable towards
phase separation, fragmenting eventually, we believe, into an ``emulsion'' of  
quasiparticle-rich droplets in a nearly quasiparticle-free matrix.\footnote{The term 
``mixed state'' may then be quite apt,
as such a state would be reminiscent of the mixed state of type II superconductors in
a magnetic field, with a difference that the droplets now carry
``electric'' rather
than magnetic fluxes.} Some implications of this picture are discussed in the conclusion.

\section{Preliminaries}
The 10-dimensional metric sourced by 
$N$ coincident extremal D3 branes in type IIB supergravity is \cite{Horowitz:1991cd} 
\be
ds^2 = \frac{1}{\sf} \left( - dt^2 + (d x^i)^2 \right) +
\sf \left( d\rho^2 + \rho^2 d\Omega_3^2 + d \D^2 + \D^2 d \phi^2  \right) \, .
\label{ds2}
\ee
The coordinates along the D3s are $t$ and $x^i$, $i=1,2,3$. The transverse coordinates
are $x^4,\ldots,x^9$, out of which we have constructed a spherical system
for $x^4,\ldots,x^7$, with radius $\rho$, and a polar system for $x^8, x^9$, with radius
$\D$. Thus, $\phi$ is equivalent to $\phi+ 2\pi$. The metric function $f$ depends only 
on $r=(\rho^2 + \D^2)^{1/2}$ and equals
\be
f(r) = 1 + \frac{R^4}{r^4} = 1 + \frac{R^4}{(\rho^2 + \D^2)^2} \, ,
\label{f}
\ee
where 
\be
R^4 = 4\pi g_s (\alpha')^2 N \, ,
\label{R4}
\ee
$g_s$ is the closed string coupling, and $1/(2\pi \alpha')$ is the string tension. 

The probe D5 wraps $x^1$ and $x^4,\ldots,x^7$ (breaking all supersymmetries). Thus,
the only spatial direction common to all branes is $x^1 \equiv x$, and the directions 
in which all branes have definite positions are $\D$ and $\phi$. The complex position 
\be
\Psi = \D e^{i\phi} = x^8 + i x^9 
\label{op}
\ee
of the D5 relative to the D3s plays the role of a superconducting order parameter.
We will also use the real position vector
\be
\X =(x^8, x^9) = (\D \cos\phi, \D\sin\phi) \, .
\label{X}
\ee

Embedding the D5 in the geometry (\ref{ds2}) means specifying $x^2, x^3, x^8, x^9$ and 
the worldvolume gauge field $A$, all as functions of the worldvolume coordinates.
In this paper, we consider only embeddings that have $x^2 = x^3 = 0$ and are
constant over the 3-sphere in (\ref{ds2}). These, then, are specified by 
\ba
\X & = & \X(t,x,\rho) \, , \label{embX} \\
A_a & = & A_a(t, x, \rho) \, , \label{embA}
\ea
where $a= t,x, \rho$.
The normal state corresponds to
\ba
\X & = & 0 \, , \label{trivX} \\
A_t & = & A_t(\rho) \, \label{trivA} ,
\ea
with all the other components of $A$ equal to zero. We refer to this as the trivial embedding.

As we will see, the trivial embedding is unstable: the D5 develops a
nontrivial
profile $\D(x,\rho)$ with characteristic magnitude $\D \sim R$. As a result, 
the near-horizon (decoupling) limit  $r \ll R$, in which the background (\ref{ds2}) approaches
the $\mbox{AdS}_5 \times \mbox{S}^5$ space, and the type IIB string theory on it becomes
dual to a conformal
field theory (CFT) \cite{Maldacena:1997re,Gubser:1998bc,Witten:1998qj}, 
cannot be taken here. This means that, in the description of the SC state, one cannot replace 
the 3/5 strings with their ground states; the entire ladder of excited string states remains. 
In a superconductor, that can be interpreted as
quasiparticles acquiring an internal structure. While there is nothing wrong with this
in principle, in practice one faces the problem of how to define, let alone use, this theory.
On the other hand, on the gravity side, the low-energy modes are described by the action of
the D5 embedded in the full D3 geometry (\ref{ds2}), and the high-energy (stringy) modes
are described by  strings connecting the branes.
As a result, many properties of the superconductor can be computed on the gravity side
even without a complete definition of the dual
quantum theory. In this paper, we consider several of these properties. They are
(i) the symmetry breaking pattern, (ii) the
quasiparticle gap (which is given by the minimal energy of the 3/5 strings), and (iii) the
free energy, computed from the action of the D5 in the geometry (\ref{ds2}).

Our theory has two U(1) symmetries: one is the phase rotation of $\Psi$, and the other
is the gauge symmetry on the D5 worldvolume; the latter has (\ref{embA}) for the gauge field.
The first U(1) is spontaneously broken by a nonzero $\D$,\footnote{
Quantum fluctuations that could conceivably restore this symmetry in a thin wire are not seen
in the leading order of the large-$N$ approximation.} but the worldvolume U(1) remains
exact.\footnote{
This makes our construction quite different from that of \cite{Ammon:2008fc}, 
in which superconductivity is related
to breaking of a worldvolume gauge symmetry.}
The corresponding conserved charge has been 
identified with the linear momentum (in units of the Fermi momentum) 
in the superconductor \cite{Khlebnikov:2012yd}, 
and we will mention one motivation for that shortly. Incidentally, this identification
implies that, to describe a disordered superconductors,
where momentum is not conserved, one will need will need a 
mechanism for breaking the worldvolume U(1).
We will touch upon this problem in the conclusion but for the rest of
the paper proceed with the clean, momentum-conserving case.

The action for the D5 consists of the DBI action 
and a Chern-Simons (CS) term 
\cite{Polchinski:Vol2}; the latter describes interaction of the D5 with the
5-form field strength sourced by the D3s. For embeddings of the form 
(\ref{embX}), (\ref{embA}), the action can be written concisely with the help of
a fictitious metric
\be
h_{ab} = \mbox{diag} (-1/f, 1/f, 1) \, ,
\label{hab}
\ee
where $f$ is the function (\ref{f}). Let us also define the quantities
\be
B^a = \half \eps^{abc} F_{bc} \, ,
\label{mag}
\ee
where $F_{ab} = \partial_a A_b - \partial_b A_a$ and $\eps^{abc}$ is the 
completely antisymmetric unit tensor ($\eps^{tx\rho} = 1$), and the cross product
\be
\X_{,a} \times \X_{,b} = x^8_{,a} x^9_{,b} - x^9_{,a} x^8_{,b} \, .
\label{cross}
\ee
Subscript commas denote partial derivatives.

In what follows, we choose $R$ as our unit of length: $R=1$.

By lowering (raising) indices with $h_{ab}$ (its inverse),
the DBI term can be written as
\be
S_{DBI} = - 2 \pi^2 \tau_5 \int dt dx d\rho \rho^3 \sf D^{1/2} \, ,
\label{DBI}
\ee
where
\be
D = 1 + \X^{,a} \X_{,a} + \half (\X^{,a} \times \X^{,b}) (\X_{,a} \times \X_{,b})
- f \left[ B^a B_a + (B^a \X_{,a})^2 \right] \, ,
\label{D}
\ee
and the CS term as
\be
S_{CS} = 2\pi^2 \tau_5 \int dt dx d\rho \eps^{abc} A_a \phi_{,b} \Pi_{,c} \, ,
\label{CS}
\ee
where
\be
\Pi(t,x,\rho)  = \frac{\rho^4}{[\rho^2 + \D^2(t,x,\rho)]^2} \, .
\label{Pi}
\ee
In (\ref{DBI}) and (\ref{CS}), 
\be
\tau_5 = \frac{1}{(2\pi)^5 g_s (\alpha')^3} 
\label{T5}
\ee
is the brane tension.

\section{Identification of the momentum components} \label{sec:comp}
Variation of $S_{CS}$ with respect to $A_a / 2\pi \alpha'$ is the conserved U(1) current 
induced on the D5 worldvolume:
\be
K^a = 4\pi^3 \alpha' \tau_5  \eps^{abc} \phi_{,b} \Pi_{,c}  = 
\frac{N}{2\pi} \eps^{abc} \phi_{,b} \Pi_{,c} \, .
\label{Ka}
\ee
The temporal component of this is the worldvolume
charge density. We see that wound configurations
of the D5 (i.e., those for which the phase gradient $\phi_{,x}$ is nonzero) carry
a charge density proportional to 
$\phi_{,x}$. This motivates the identification of the total induced charge,
\be
P_s = \int dx d\rho K^t \, ,
\label{Ps}
\ee
with the momentum (in units of the Fermi momentum $k_F$)
carried by the SC component. 
(The fact that the coefficient in the relation between $P_s$
and the momentum is exactly unity can be established by observing that a single 3/5 string
carries unit charge and corresponds to a quasiparticle that carries $\pm k_F$ of momentum.)

It may be more familiar (see, for example,  \cite{Ammon:2008fc}) to have 
the variational derivative with respect to $A_t$ correspond to the number density 
in the dual theory, and 
the derivative with respect to $A_x$ to the current density. Here, we relate
$\delta S_{CS} / \delta A_t$ to the momentum, which is more like current than charge. To see
how this transmutation of charge into current comes about in the dual theory, 
consider the way the original
electron operators are packaged into the Dirac fermions representing the ground states of
the 3/5 strings. The fermions, $\chi_n$, are (1+1)-dimensional but carry a Chan-Paton index 
$n =1,\ldots, N$. In the dual superconductor, $n$ corresponds to the conductance channel,
i.e., the transverse wavefunction, occupied by the fermion \cite{Khlebnikov:2012yd}
(thus, a larger $N$ means a thicker sample). Each $\chi_n$  
is two-component, and the bilinear that couples to the order
parameter (\ref{op}) is ${\cal O} = \sum_n \bar{\chi}_n ( 1+ \gamma^5) \chi_n$.
Choose the representation where the Dirac $\gamma^0$ is the
Pauli $\sigma_1$, and $\gamma_5$ is $\sigma_3$. Then, for the bilinear ${\cal O}$ to represent
the SC pairing channel, the upper component of each $\chi$ 
(we omit the subscript $n$ now) must be the
annihilation operator of a right-moving electron, $a_R$, and the lower component the 
creation operator of a left-moving one, $b_L^\dagger$. As a result,
the normally ordered ``charge'' density 
$\bar{\chi} \gamma^0 \chi$, which couples to $A_t$, is 
proportional to $a_R^\dagger a_R - b_L^\dagger b_L$, i.e., the momentum
density (in units of $k_F$) in the dual theory, and the ``current'' 
density $\bar{\chi} \gamma^1 \chi$, which couples to
$A_x$, to the number density.

A useful expression for $P_s$ can be obtained from the Gauss law, the temporal
component of the equation of motion for $A_a$. The equation reads
\be
- \eps^{abc} \partial_b \left( \rho^3 \sf H_c \right) = \frac{2\pi}{N} K^a \, ,
\label{eqmA}
\ee
where
\be
H_c = - \dpar{\sqrt{D}}{B^c} = 
\frac{f}{\sqrt{D}} \left[ B_c + (B^b \X_{,b}) \cdot \X_{,c} \right] \, .
\label{Hc}
\ee
From now on we assume that the $x$ direction is a circle (of length $L$). Then,
setting $a =t$ in (\ref{eqmA}) and integrating over $x$ and $\rho$ gives
\be
\int dx d\rho K^t = P(\infty) - P(0) \, ,
\label{gauss}
\ee
where
\be
P(\rho) = \frac{N}{2\pi} \int dx \rho^3 \sf H_x \, .
\label{Prho}
\ee

$P(\rho)$ is the flux of the ``electric'' induction through a surface of constant $\rho$. 
The value
$P(0)$ is nonzero only if the D5 crosses the horizon, $r=0$, of the geometry (\ref{ds2}),
which solutions considered in this paper do. One can visualize it as an effect of 
3/5 strings that have fallen through the horizon and pulled the D5 with them. In this
picture, $P(0)$ is the total worldvolume charge carried by these strings.\footnote{
We wish to stress here that,
unlike in descriptions of supercurrent that use a bulk gauge field
\cite{Gubser:2008px,Hartnoll:2008vx,Nakamura:2009tf}, 
in our case the ``electric'' field lives only on the 
D5 worldvolume.}
Alternatively,
one could think of $P(0)$ simply as an additional variable characterizing the boundary 
of the D5 at $r=0$. The interpretation of it as a charge ``behind the horizon,'' however,
is helpful in understanding that this variable is dynamical (as we discuss in more detail
in Sec.~\ref{sec:fren}): if additional quasiparticles (3/5 strings) are
produced by unwinding the supercurrent and fall through the horizon, both 
$P_s$ and $P(0)$ change;
only the total, $P(\infty)$, is conserved. According to our 
interpretation of the charge as momentum, $P(\infty)$ is the total momentum of
electrons in the wire. Comparing (\ref{gauss}) with (\ref{Ps}), we see that
$P_n = P(0)$ can be identified with the momentum of the normal component: it
adds to the supercurrent momentum $P_s$ for the total 
of $P \equiv P(\infty)$, cf. eq.~(\ref{Pn}).

Eq.~(\ref{eqmA}) has a class of solutions (first integrals) of the form
\be
\rho^3 \sf H_a = \phi_{,a} (\Pi - 1) + J_a \, ,
\label{fint}
\ee
where $J_a$ are integration constants. These solutions are not the most general, 
as one can always 
add a gradient to the right-hand side of (\ref{fint}) without affecting the curl 
in (\ref{eqmA}), but they
will be sufficient for our purposes. In fact, we will restrict the class of solutions 
even further---to those for which the only nonzero constant is $J_x$, and we will use 
a special notation for it: 
\ba
& & J_x \equiv \Jinf \, , \label{Jx} \\
& & J_t = J_\rho = 0 \, . \label{Jtrho}
\ea
As we will see shortly, $\Jinf$ corresponds to the total (SC plus normal) momentum density in 
the wire. One may suspect that, similarly, a nonzero $J_t$ would describe
variations in the fermion number density (that is, in the Fermi momentum) but,
as (\ref{Jtrho}) implies,
that will not be pursued here.

Using (\ref{fint}), eq.~(\ref{Prho}) can be written as
\be
P(\rho) = \frac{N}{2\pi} \int dx \left[ \phi_{,x} (\Pi - 1) + \Jinf \right] \, .
\label{Prho2}
\ee
For all solutions considered in this paper $\Pi(x,\infty) = 1$. This is a consequence
of the boundary condition 
\be
\D \to 0 \hspace{1em} \mbox{at} \hspace{1em} \rho \to \infty \, ,
\label{inf}
\ee
which means that there is no symmetry-breaking source, i.e., 
the U(1) that rotates the phase of $\Psi$ is broken spontaneously rather than 
explicitly. Then,
at large $\rho$, the integrand on right-hand side of (\ref{Prho2})
is simply $\Jinf$. According to our earlier 
interpretation
of $P(\infty)$, this means that $\Jinf$ is, up to a factor of $N/2\pi$, 
the linear density of the total
momentum.\footnote{This suggests a generalization (which we do not pursue here)---a 
$J_\infty$ depending arbitrarily on $x$. This does not affect the curl in eq.~(\ref{eqmA}),
so the equation is still satisfied. An $x$-dependent $\Jinf$ may describe, for instance,
supplying extra current to segments of the wire by connecting them to
external leads.} Similarly, sending $\rho\to 0$ allows us to identify 
\be
J_s(x) = \phi_{,x}(x,0) [1 - \Pi(x,0) ] 
\label{Js}
\ee
as the momentum density of the supercurrent (up to the same factor)
and the difference $\Jinf - J_s$ as that of the normal component.

\section{Instability of the normal state} \label{sec:inst}
Consider linearized theory near the trivial embedding (\ref{trivX}), (\ref{trivA}). 
To the linear order in $\X$,
\be
H_c = \frac{f_0 B_c}{(1 - f_0 B^a B_a)^{1/2}} 
\label{Hlin}
\ee
where
\be
f_0 = 1 + \frac{1}{\rho^4} \, .
\label{f0}
\ee
To this order, subject to the conditions (\ref{Jtrho}), there are no sources 
in (\ref{fint})
for $H_t$ and $H_\rho$, while for $H_x$ the only source is $\Jinf$. As a result, 
$B^a$ are unchanged from the zeroth order, namely, $B^t = B^\rho = 0$ and
\be
B^x(\rho) = - F_{t\rho}(\rho) = \frac{\Jinf}{C_0^{1/2}(\rho)} \, ,
\label{Bxlin}
\ee
where
\be
C_0 (\rho)= \rho^6 f_0(\rho) + \Jinf^2 \, .
\label{C0}
\ee

Upon substituting (\ref{Bxlin}), the linearized equation for $\X$, written in terms of 
the complexified coordinate (\ref{op}), reads
\be
\frac{1}{\sqrt{C_0}} \dpar{}{\rho} \left( \sqrt{C_0} \Psi_{,\rho} \right)
+ \frac{2 \Psi}{C_0} - f_0 \ddot{\Psi} + \frac{\rho^6 f_0^2}{C_0} \Psi_{,xx} 
- i \frac{4 J_\infty}{\rho^2 C_0}  \Psi_{,x} = 0 \, .
\label{lin}
\ee
The general solution is a superposition of plane waves
\be
\Psi(t,x,\rho) = \D(\rho) e^{-i\omega t + i qx} \, .
\label{pwave}
\ee
Here $q$ is real but $\omega$ can be complex. For $q\neq 0$, eq.~(\ref{pwave}) describes 
a D5 uniformly
would about the D3s, the total of $W = qL / 2\pi$ times. As we have already noted, in
our interpretation, such wound states describe supercurrent. 

We are interested in
unstable modes---those $\D(\rho)$ for which $\omega$ has a positive imaginary part.
Suppose $\Jinf > 0$. Then,
at small $\rho$, unstable modes behave as $\D \sim \rho \exp(i\omega / \rho)$, i.e., vanish
exponentially. They also vanish at $\rho\to \infty$. 
The equation for $\D$ obtained by substituting (\ref{pwave}) in
(\ref{lin}) has the form of a Schr\"{o}dinger equation, and the boundary conditions 
just established  mean that the unstable modes are its bound states. Asymptotically, 
at small $\rho$, the equation is
\be
\D_{,\rho\rho} + \frac{\omega^2}{\rho^4} \D + \frac{\alpha}{\rho^2} \D = 0 \, ,
\label{fall}
\ee
where
\be
\alpha = 4 q / \Jinf - (q / \Jinf)^2 \, .
\label{alpha}
\ee
A change of the independent variable to $z = 1/\rho$ shows that this is a  
``fall to the center'' problem. It becomes supercritical for $\alpha > \frac{1}{4}$,
which means that for these $\alpha$ the full 
Schr\"{o}dinger problem has an infinite number of bound 
states \cite{Landau&Lifshitz}. In our case, such $\alpha$ exist for any $\Jinf > 0$,
with the maximum $\alpha = 4$ reached at $q/\Jinf =2$.
We conclude that, for any $\Jinf > 0$, the normal state is unstable.
The instability band is at least as broad as the supercritical range
\be
2 - \frac{\sqrt{15}}{2} < \frac{q}{\Jinf} < 2 + \frac{\sqrt{15}}{2} \, ,
\label{range}
\ee
but may actually be broader, since it will include also those $q$ for which there is
only a finite number of unstable modes.

\begin{figure}
\begin{center}
\includegraphics[width=4in]{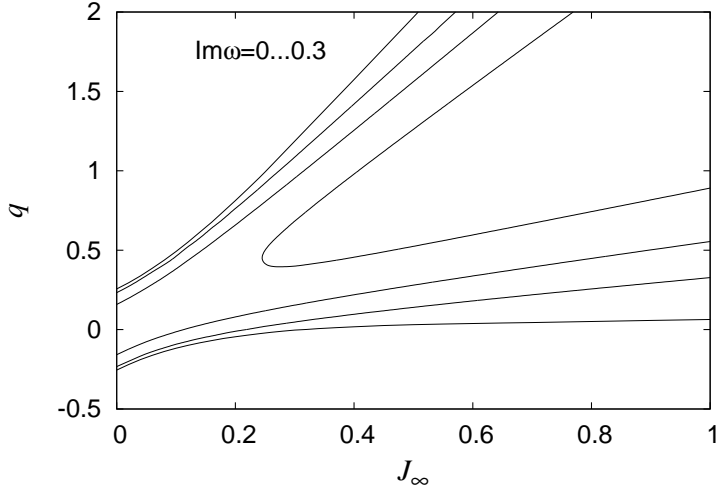}
\end{center}                                              
\caption{Instability chart of eq.~(\ref{lin}) after substitution (\ref{pwave}).
The lines are the level contours of $\mbox{Im} \omega$ as it is increased in increments of
0.1. The outermost pair of lines corresponds to $\mbox{Im} \omega = 0$ and forms 
the boundary of the instability band.
}                                              
\label{fig:cont}                                                                       
\end{figure}

A numerical solution to the full eq.~(\ref{lin}) produces the instability chart shown in
Fig.~\ref{fig:cont}. Note that, for comparatively small $\Jinf$, the instability 
band includes the value $q = 0$, even though the latter is not in the supercritical range 
(\ref{range}), but for large  $\Jinf$, the instability occurs
only for modes with $q$ above a certain nonzero minimum: the emerging
SC state necessarily has a supercurrent.\footnote{
For a loop of a finite length $L$, the values of $q$ are quantized, and for a short loop 
it is possible that, at a given $\Jinf$, all the allowed values of $q$ fall outside 
the instability band. In this case, there will be a curious reentrant behavior, when
the normal state, linearly stable at that $\Jinf$, becomes linearly
unstable again at a larger one.}

\section{Uniform mixed states}
The next question is what is the final state that the instability leads to. 
Natural first tries are the simplest configurations---those that are time-independent 
and uniform. By the latter we mean that $\D$ is independent of $x$, while 
$\phi$ winds along $x$ uniformly:
\ba
&  & \D = \D(\rho) \, , \\
&  & \phi = \phi(x) = q x \;  .
\label{uni}
\ea
For such configurations, the matrix 
\be
M_{cb} = h_{cb} +  \X_{,c} \cdot \X_{,b} 
\label{M}
\ee
multiplying $B^b$ in (\ref{Hc}) is diagonal. Expressing $H_a$ from (\ref{fint}), subject
to the conditions (\ref{Jtrho}), and $B^a$ from (\ref{Hc}), we find
\ba
& & B^t = B^\rho = 0 \, \label{Btrho} \\
& & B^x(\rho) = - F_{t\rho}(\rho) = \frac{J(\rho)}{C^{1/2}(\rho)} (1 +\Drho^2)^{1/2} \, ,
\label{Bx}
\ea
where the various functions (of $\rho$ only) are given by
\ba
J(\rho) & = & q [ \Pi(\rho) - 1 ] + \Jinf \, , \\
C(\rho) & = & \rho^6 f_\D(\rho) \left[ 1 + q^2 \D^2(\rho) f_\D(\rho) \right] 
+ J^2(\rho) \, , \\
f_\D(\rho) & = & 1 + \frac{1}{[\rho^2 + \D^2(\rho)]^2} \, .
\ea
These equations are the 
nonlinear counterparts to eqs.~(\ref{f0}), (\ref{Bxlin}), (\ref{C0}) of the linear theory.

The equation of motion for $\D(\rho)$, obtained by varying the action 
and substituting (\ref{Btrho}) and (\ref{Bx}), is
\be
\frac{d}{d\rho} \frac{\Drho \sqrt{C}}{(1 + \Drho^2)^{1/2}}
=  (1 + \Drho^2)^{1/2}  \dpar{\sqrt{C}}{\D} \, .
\label{eqm}
\ee
The boundary condition at infinity is (\ref{inf}). We now proceed to establish the condition
at $\rho \to 0$.

The first part of the argument is standard (for an application to a different system,
see for example \cite{Kobayashi:2006sb}). Recall from Sec.~\ref{sec:comp}, eq.~(\ref{Prho2}),
that $J(\rho)$ represents the flux of the worldvolume
electric field through a surface
of constant $\rho$. If $\D(0) \neq 0$, the D5 closes off at a finite distance from
the D3s. In this case, we must have $J(0) = 0$; otherwise, the lines of the field
have nowhere to end. More formally, for $J(0) \neq 0$, eq.~(\ref{Bx})
predicts $F_{t\rho}(0) \neq 0$, meaning that the gauge field is not smooth. The only
way to accommodate $J(0) \neq 0$ therefore is to have $\D (0) = 0$. Then, the D5 crosses the
horizon, and the flux at $\rho = 0$ can be ascribed to charges behind the horizon,
as discussed in Sec.~\ref{sec:comp}.
According to the interpretation of the fluxes there, for a uniform
solution, $J(0)$ is the momentum density of the normal component. As we are interested
here specifically in solutions for which that is nonzero, we postulate\footnote{
The supercurrent-only solutions of \cite{Khlebnikov:2012yd}, in contrast, have
$J(0) = 0$ and $\D(0) \neq 0$.}
\be
\D(0) = 0 \, .
\label{bc0}
\ee
Eq.~(\ref{bc0}) implies that the shortest strings connecting the D5 to the D3s are of 
length zero, i.e., the superconductor is gapless, which is
consistent with the presence of a normal component.

The second part of the argument seeks to establish the manner in which $\D(\rho)$ vanishes
at $\rho\to 0$. The only type of solutions we have been able to find are those for which
that happens slower than linearly, with $\D$ maintaining its sign
(for definiteness, positive) at small nonzero $\rho$:
\be
\rho^{-1} \D(\rho) \to \infty \hspace{1em} \mbox{at} \hspace{1em} \rho \to 0 \, .
\label{slower}
\ee 
The precise asymptotics is discussed in Sec.~\ref{sec:hor}. 

Under the condition (\ref{slower}), $\Pi(0)$ in eq.~(\ref{Js}) is zero 
(and there is no longer a dependence on $x$ as the solution is uniform), so according
to that equation the momentum density of the supercurrent is
\be
J_s = q \, .
\label{Js2}
\ee
This is the same expression as obtains for the supercurrent-only solutions, 
cf. eq.~(\ref{Ps}). It is
as if each electron in the wire contributes momentum $q/2$ to the superflow.\footnote{
The total number of electrons in a wire with $N$ channels 
is $N k_F L/\pi$ (we define channels so that 
each contains only one projection of spin). Dividing
$N W k_F$ by this number gives $q/2$ per electron.} The reason why this applies
even in the presence of a normal component is that, under our present approximations,
the number of ``normal electrons'' is much smaller than the total number: with
the length scale $R$ restored, the former is $P_n \sim N L / R$, while the
latter is of order $N L k_F$. In other words, although $P_s$ and $P_n$ may be 
comparable,
the first of these is due to a large number of ``superconducting electrons'' 
each contributing the small momentum
$q/2$, while the second is due to a small number of ``normal electrons'' each carrying
the large momentum $k_F$.

\section{Near-horizon limit and numerical solutions} \label{sec:hor}

Consider the limit of eq.~(\ref{eqm}) at $\rho\to 0$. Recalling the condition
(\ref{slower}), we can expand (\ref{eqm}) in $\eps = \rho / \D$. 
For a mixed-state solution, we may assume, without loss of generality, that $J(0) > 0$ 
and $q \neq 0$. Then,
\be
\sqrt{C} =  J(\rho) + O(\eps^6) =  J(0) + \frac{q \rho^4}{\D^4} + O(\eps^6) \, .
\label{sqrtC}
\ee
Assuming that $\Drho$ is of order $1/\eps$, we find that, to the leading
order in $\eps$, the limiting form of (\ref{eqm}) is 
\be
j \frac{d}{d\rho} \frac{1}{ \Drho^2} = \frac{8 \rho^4 \Drho}{\D^5} \, .
\label{lim}
\ee
The parameter
\be
j \equiv J(0) / q = (\Jinf - q) / q
\label{j}
\ee
is the ratio of the momentum densities of the normal and SC components.

Eq.~(\ref{lim}) is scale-invariant:
if $\D(\rho)$ is a solution, then so is
\be
\widetilde{\D}(\rho) = c \D(c^{-1} \rho) \, ,
\label{scaled}
\ee
where $c$ is any positive constant. We can think of $c$ as a shooting parameter, which
we may hope to adjust so as to obtain a solution to the full eq.~(\ref{eqm}) with the
correct asymptotics (\ref{inf}) at infinity.
Indeed, this is precisely how we are going to search for solutions
to eq.~(\ref{eqm}) numerically.

The form of (\ref{lim}) suggests that it is advantageous to view $\rho$ as a function
of $\D$, rather than $\D$ as a function of $\rho$. Then, (\ref{lim}) can be rewritten as
\be
j \frac{d}{d\D} \rho_{,\D}^2 = 8 \frac{\rho^4}{\D^5} \, .
\label{rhoD}
\ee
We are looking at this in the limit $\D\to 0$, with the boundary condition
$\rho(0) = 0$. The substitution
\be
\rho(\D) = g(z) \D \, ,
\label{g}
\ee
where
\be
z = \ln ( \D_0 / \D) \, ,
\label{z}
\ee
brings (\ref{rhoD}) to the form
\be
g g' - (g')^2 - g g'' + g' g'' = - \frac{4g^4}{j}  \, .
\label{eqg}
\ee
$\D_0$ in (\ref{z}) is an arbitrary constant, playing the same role as $c$ in 
(\ref{scaled}). Note that a small $\Delta$ means a large $z$.

Eq.~(\ref{eqg}), at large $z$, is suitable for an 
application of the WKB approximation. That amounts to
ordering the terms on the left-hand side according
to the number of derivatives: the more derivatives, the smaller the term. To the
leading order, only the first term matters, and we find
\be
g^2(z)_{\rm LO} = \frac{j}{8 (z + z_0)} \, ,
\label{LO}
\ee
where $z_0$ is an integration constant. $z_0$ can be absorbed by a redefinition of 
$\D_0$ and, in any case, is immaterial to the leading order. From (\ref{LO}), we
conclude that a solution with the postulated
asymptotics exists only for $j > 0$. Referring now to eq.~(\ref{j}), we see that
$j > 0$ implies
that $J(0)$, $q$, and $\Jinf$ are all of the same sign---which, by the assumption
we have made regarding $J(0)$, is positive. In other words, we may expect to find a solution
of the requisite form only for 
\be
0 < q < \Jinf \, .
\label{qcond}
\ee
This stands to reason: the condition (\ref{qcond}) means that the SC and normal
components flow in the same direction.

Numerically, given $\Jinf$ and $q$,
we choose a small $\D$ and compute $\rho$
and $\rho_{,\D}$ from (\ref{g}), with $g$ given by (\ref{LO}) and $\D_0$ a parameter. 
We then use 
this $\D$ and the computed $\Drho = 1/ \rho_{,\D}$ as boundary conditions 
for the full eq.~(\ref{eqm}) and
look for $\D_0$ such that the solution satisfies also the condition (\ref{inf}).

\begin{figure}
\begin{center}
\includegraphics[width=4in]{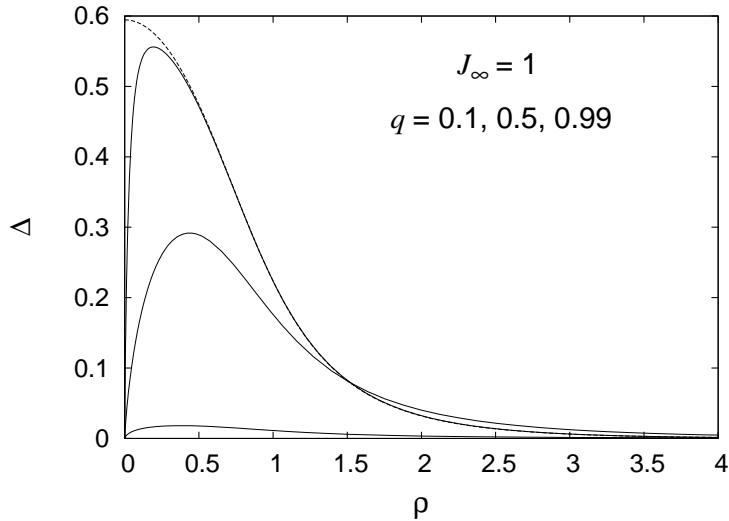}
\end{center}                                              
\caption{D5 profiles for mixed-state solutions with 
total momentum density $\Jinf =1$ and various values
of the supercurrent momentum density $q$. Larger $q$ correspond to larger peak values of
$\D(\rho)$. The dashed line is the supercurrent-only solution of \cite{Khlebnikov:2012yd}
with $q=1$.
}                                              
\label{fig:sol_Jinf=1}                                                                       
\end{figure}

Using this algorithm, 
we find that the upper bound in (\ref{qcond}) is saturated, in the sense that 
there are solutions with $q$ very close to $\Jinf$, but the lower one in general
is not: a more precise condition is
\be
\qmin < q < \Jinf \, ,
\label{qcond2}
\ee
where $\qmin$ is the larger of zero and the lower instability bound of 
Sec.~\ref{sec:inst} (the lowest curve in Fig.~\ref{fig:cont}). 
Numerically, $\qmin$ departs from zero at $\Jinf = 0.32$. At the instability
bound, the solution merges into the normal-state solution $\D\equiv 0$. This is
illustrated in Fig.~\ref{fig:sol_Jinf=1}. On the other hand, as $q$ approaches $\Jinf$,
the solution matches, except at the smallest $\rho$, the gapped supercurrent-only 
solution of \cite{Khlebnikov:2012yd}. In this way---for those 
$\Jinf$ for which $\qmin\neq 0$---a family
of mixed-state solutions with different values of 
$q$ can be thought to interpolate between
the normal state (in which $q$ is undefined) and the supercurrent-only state 
with $q = \Jinf$.

\section{Non-convexity of the free energy} \label{sec:fren}
Eq.~(\ref{eqm}) is the condition of local extremum, with respect to 
$\D(\rho)$, for the functional
\be
F = \int_0^\infty d\rho \left[ \sqrt{C} ( 1 + \Drho^2)^{1/2} - \rho^3 \sfz \right] \, ,
\label{fren}
\ee
where $f_0$ is given by (\ref{f0}).
This identifies $F$, up to an overall normalization, as the free energy density.
The last term under the integral 
does not depend on $\D$ and so does not contribute to (\ref{eqm}); its role is
to make the integral convergent at the upper limit.

In a clean (disorder-free) conductor, a phase slip changes the momentum carried 
by the supercurrent without changing the total
momentum. In our case, the former is represented by $q$ and the latter by $\Jinf$,
so it makes sense to consider $F$ as a function of $q$ at fixed $\Jinf$. The minimum
of this function will be a candidate ground state---not necessarily the true one, as
the procedure applies to uniform states only. 

\begin{figure}
\begin{center}
\includegraphics[width=4in]{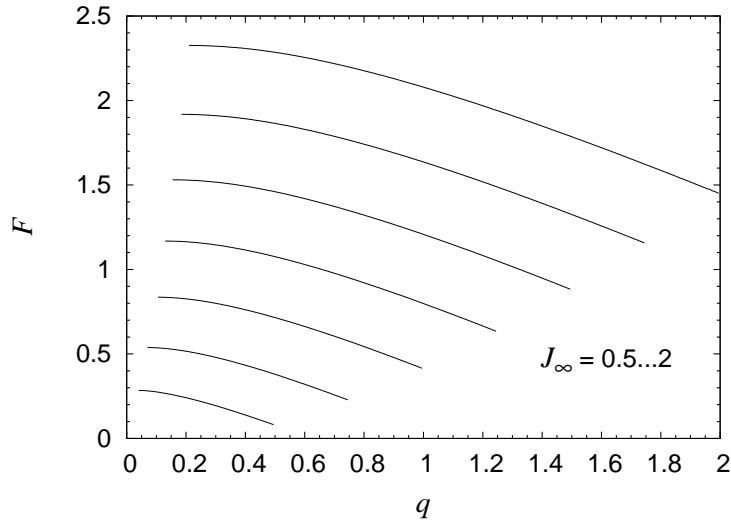}
\end{center}                                              
\caption{Free energy of a uniform mixed state as a function of $q$. $\Jinf$
increases bottom to top in increments of 0.25.
}                                              
\label{fig:fren}                                                                       
\end{figure}

Numerically computed $F(q)$ curves for several values of $\Jinf$ are shown in 
Fig.~\ref{fig:fren}.
The ends of each curve correspond to the endpoints of the interval (\ref{qcond2}).
At the right end, $q \to \Jinf$, the free energy is the same as for the gapped
solution to which the mixed-state solution converges pointwise 
(cf. Fig.~\ref{fig:sol_Jinf=1}). 

We see that---among uniform states with a given $\Jinf$---the gapped, supercurrent-only
state has the lowest free energy. Based on that, one might suppose that this is the state
that the system will always evolve to. With the aid of Fig.~\ref{fig:sol_Jinf=1}, one could 
visualize such an evolution as the D5 peeling itself off
the horizon, to form the state represented by the dashed line. From
the earlier work \cite{Khlebnikov:2012yd} we know, however, that above a certain value
of $\Jinf$,
\be
\Jinf > \Jinf^{(m)} \, ,
\label{Jm}
\ee 
the gapped state is unstable to decay by phase slips, which
is accompanied by production of quasiparticles.\footnote{For production of
$N$ well separated quasiparticles in a long wire, 
$\Jinf^{(m)}$ is about 0.57 \cite{Khlebnikov:2012yd}. 
The threshold may be lower for production of a bound state.}
What we learn here, then, is not that the gapped state is always stable, but rather
that the quasiparticles that are produced by its decay cannot be described by a uniform
solution.

This conclusion is supported by the 
observation that none of the curves in Fig.~\ref{fig:fren} 
is convex. That means that, in a long wire, 
a uniform state with $q < \Jinf$ is unstable with respect to phase 
separation---fragmentation into regions
with different winding number densities.\footnote{This instability is distinct from 
the ones leading to spatially modulated currents \cite{Nakamura:2009tf,Amado:2013aea}
in models using a bulk U(1) field. In our case, the system has fewer translational
directions, and the total current remains uniform. It
is the partition of the total into superconducting and normal components that becomes
$x$-dependent.}
Unlike the instability of the gapped state,
this one does not rely on phase slips: it occurs even at fixed total winding,
equal to that of the initial uniform state. The form of the free-energy curves 
(with the absolute minimum reached at $q = \Jinf$) suggests that, upon phase
separation, there will be regions 
with $q\approx \Jinf$, which are almost quasiparticle-free, while quasiparticles are
concentrated in droplets dispersed among these regions---an ``emulsion.''

We find plausible the following hypothesis about the nature of the quasiparticle-rich
droplets: they are ``baryons,'' each made of $N$ D3/D5 strings. Recall that these strings
carry color with respect to the $SU(N)$ gauge group that lives on the D3s' worldvolume.
The baryon is colorless.\footnote{It is similar to the baryon vertex described
in \cite{Witten:1998xy,Gross:1998gk}, 
except that in our case the $N$ strings connect the D5 to the D3s, 
rather than to the boundary of an AdS space.} 
The complete antisymmetry of its wavefunction
with respect to color means that there is a quasiparticle in each of the $N$ conductance
channels of the wire.

\section{Conclusion}

In the present work, we have aimed to understand the nature of the mixed SC-normal state
that forms in a strongly coupled thin superconductor at currents above a certain $I_m$ but
well below the depairing current $I_{dep}$. This has been done here in the context of the 
same D3/D5 system as we used in \cite{Khlebnikov:2012yd}.
The momentum carried by the supercurrent is represented by the flux of the worldvolume
gauge field induced on the D5, as the latter winds around the D3s, and the momentum of 
the normal component by the flux due to charges behind the horizon. Natural first guesses for the mixed state
are uniform solutions, in which both these fluxes are uniform in $x$ (the coordinate along the wire) and, in
the leading large $N$ limit, entirely classical. We have argued that such solutions exist but are unstable
against fragmentation, leading eventually, we believe, 
to a non-uniform ground state---an ``emulsion'' of quasiparticle-rich droplets in a nearly 
quasiparticle-free matrix.

The non-uniform 
mixed state hypothesized here is similar to the mixed state of type II superconductors, 
with the droplets seen as ``electric'' analogs of the magnetic flux lines,
and the winding density $q_m$, at which production of quasiparticles becomes energetically 
favorable, as the counterpart of the lower critical field $H_{c1}$. 
This analogy leads us to speculate further on the properties of such a state, in particular,
on the role of disorder. All quasiparticles produced by phase slips carry
the same momentum (either $k_F$ or $-k_F$). One may wonder, then, 
if in the presence of disorder, when momentum is no longer conserved, 
the quasiparticles will not simply disappear, and the wire will not
revert to the purely SC state. The analogy with a type II superconductor makes us think that
this is unlikely. In that case, the presence of flux lines in the sample is not a result
of any conservation law (they can enter and leave the sample through the boundary)
but a consequence of the energetics: 
the difference $H- H_{c1}$ (in our case, $q - q_m$) plays the
role of a chemical potential for a flux line (in our case, a quasiparticle).

Guided by the same analogy, one may contemplate a periodic array of 
droplets---an analog of 
the Abrikosov flux lattice. One may wonder if there are classical solutions of dual
gravity capable of describing such an array.

The author would like to thank A. Kamenev for discussions. This work was supported in
part by the U.S. Department of Energy grant \protect{DE-SC0007884}.

\end{document}